\documentclass[a4paper,10pt]{llncs}
\usepackage[utf8]{inputenc}
\usepackage[margin=3cm]{geometry}
\usepackage{float}
\usepackage{url}

\usepackage{color}
\usepackage{amsmath}
\newcounter{todocnt}
\begin{document}
\title{UNIMIB at TREC 2021 \\ Clinical Trials Track}
\author{Georgios Peikos \and Oscar Espitia \and 
Gabriella Pasi}
\institute{Information and Knowledge Representation, Retrieval, and Reasoning (IKR3) Lab,\\
	Department of Informatics, Systems, and Communication (DISCo),\\
	University of Milano-Bicocca, Milan, Italy\\
\email{\{georgios.peikos, oscar.espitiamendoza, gabriella.pasi\}@unimib.it}}
\date{October 2020}

\maketitle
\begin{abstract}
This contribution summarizes the participation of the UNIMIB team to the TREC 2021 Clinical Trials Track. We have investigated the effect of different query representations combined with several retrieval models on the retrieval performance. First, we have implemented a neural re-ranking approach to study the effectiveness of dense text representations. Additionally, we have investigated the effectiveness of a novel decision-theoretic model for relevance estimation. Finally, both of the above relevance models have been compared with standard retrieval approaches.
In particular, we combined a keyword extraction method with a standard retrieval process based on the BM25 model and a decision-theoretic relevance model that exploits the characteristics of this particular search task. The obtained results show that the proposed keyword extraction method improves 84\% of the queries over the TREC's median NDCG@10 measure when combined with either traditional or decision-theoretic relevance models. Moreover, regarding RPEC@10, the employed decision-theoretic model improves 85\% of the queries over the reported TREC's median value.

\end{abstract}

\section{Introduction}
\label{sec:prop_intro}
The TREC 2021 Clinical Trials Track\footnote{\url{http://www.trec-cds.org/2021.html}} focuses on the task of retrieving clinical trials that are eligible for a particular patient. Specifically, patient-related information is provided as a query consisting of a synthetic case in the form of an admission note. Each query describes a patient by means of certain conditions and observations, and, in general, queries are longer than those in traditional ad-hoc retrieval tasks. Moreover, clinical trials, i.e., the considered document collection, are multi-fielded documents, where a core aspect is constituted by the fields that specify the inclusion and exclusion criteria. Those are not all-inclusive statements about the trial to the point that other trial information can be ignored, but they are essential aspects for defining a trial's eligibility (relevance) to a given patient. 

This report outlines our team's (UNIMIB) submission to TREC 2021 Clinical Trials Track, it discusses the experimental setup, and compares our results to the reported TREC's median performance. In particular, we examined the effect of different query representations for retrieving eligible clinical trials. Moreover, we combined those different query representations with several relevance models, one of which is a novel decision-theoretic model that leverages the peculiarities of this search task in the relevance estimation.  

\section{Methodology}
\label{sec:methodology}
This section describes the three components of the submitted runs, i.e., the query and document representations, along with the employed relevance models. The first one is the BM25 model that leverages an ad-hoc query representation. In addition, the second relevance model is a decision-theoretic model whose properties are based on the peculiarities of this particular search task; finally, a neural network-based model in a re-ranking setup has been employed.

\subsection{Query representation}
\label{sec:Keywords}
The provided queries consist of synthetic patient cases created by medical experts in the form of admission notes. Previous studies~\cite{DBLP:conf/sigir/KoopmanCZ17,DBLP:conf/sigir/KoopmanZ16} have investigated the effect of different query representations on the system's performance. Their findings suggested that using the whole verbose query is not as effective as using its ad-hoc representation. Therefore, given a query that is a verbose representation of a synthetic patient case, we extracted keywords that better represent its content. 

To this aim, we have employed EmbedRank++ proposed in~\cite{DBLP:conf/conll/Bennani-SmiresM18}, which is a fast, unsupervised, and collection-independent method for extracting key phrases and keywords; these advantages make this method suitable for extracting keywords from verbose queries. Firstly, given a query, we estimated its embedding representation, and then we estimated the embedding representation of every of its tokens, both unigrams and bigrams. Having done that, we selected the keyword or key phrase that is most similar to the query by measuring their cosine similarity. Then, by employing the maximal marginal relevance (MMR) formula, as introduced in \cite{DBLP:conf/conll/Bennani-SmiresM18}, we iteratively selected new keywords or key phrases similar to the query and different from the already extracted keywords. 
The employed keyword extraction approach requires the setting of two parameters; the total number of extracted keywords, and a parameter $\lambda$ that balances the trade-off between similarity to the query and diversity between the extracted keywords.
Concluding, by using this tool, we were able to extract those keywords that better represent each query, and at the same time, avoid redundant terms. 

For the submitted runs, we have used two different query representations, i.e., the original verbose query \textit{Qd}, provided by the organizers, and an ad-hoc representation that contains the extracted keywords, \textit{Qk}.

\subsection{Document Representation}
\label{sec:Indexing}
In the context of this particular search task, as pointed out in  Section\ref{sec:prop_intro}, several document sections have a different effect on the relevance of the whole document to the query. Specifically, without underestimating the contribution of the other document sections, the parts that refer to a trial's inclusion and exclusion criteria in the eligibility section are of great importance. 

Besides the eligibility section, other important document sections are its title, description, summary, and studied condition. 
Consequently, three of our submitted runs leverage an index created by combining the eligibility section with all the document sections mentioned above; throughout this paper, this index will refer to as \textit{I\_comb}. For the other two runs the index was created for only the title and summary
with inclusion criteria when it is specified, and successfully extracted, this index will be referred to as \textit{I\_comb*}.

Two additional indices have been created for each document, namely \textit{I\_in} which indexes only the trial's inclusion criteria, and \textit{I\_ex} which indexes only the trial's exclusion criteria.
For their extraction, several regex rules that leverage their semi-structured nature have been used.
In the cases where the extraction of the inclusion and exclusion criteria was not feasible, the whole eligibility section has been indexed for both indices, i.e., the \textit{I\_in} and \textit{I\_ex}. 
In addition to those two indices, a third one has been created. Here, the title, description, studied condition, and the summary sections were combined in a single text and indexed; this index will be referred to as \textit{I\_main}. 

In summary, the first, fourth and fifth submitted run uses the combined document representations \textit{I\_comb} and \textit{I\_comb*} to estimate the relevance of a clinical trial, while the relevance decision for the second and third submissions rely on the \textit{I\_in}, the \textit{I\_ex} and the \textit{I\_main} indices, and will be further analyzed in section~\ref{sec:dt}

\subsection{Retrieval Models}
\label{sec:retrieval}
Given the various query and document representations, we have combined them with several relevance models.
The considered models are the traditional BM25 model, a neural network-based model in a re-ranking setup, trained on previous clinical trials collections, and a
decision-theoretic relevance model, which we have introduced for this particular search task. Specifically, the decision-theoretic relevance model is based on the Technique for Order of Preference by Similarity to Ideal Solution (TOPSIS)~\cite{DBLP:books/sp/HwangY81}.

\subsection{Standard Approach}
A traditional retrieval approach has been employed that leverages the Qk representation along with the I\_comb index and the BM25 model to estimate an overall relevance value and rank the documents. 

\subsection{Decision-Theoretic Approach}
\label{sec:dt}
As stated previously, the trial's inclusion and exclusion criteria have a significant role in the relevance (eligibility) of a trial to a particular patient. Specifically, we argue that the semantic similarity of a patient health record, in our case the query, to a trial's inclusion criteria contributes positively to the trial's overall relevance. In contrast, the semantic similarity to its exclusion criteria contributes negatively. In addition, we argue that the semantic similarity of a query to the remaining trial's sections, i.e., its title, description, summary, studied condition, also contributes positively to its overall relevance. These are task-specific characteristics that standard retrieval approaches usually do not consider, while a decision-theoretic model allows us to incorporate them in the relevance estimation.

In detail, to estimate the overall trial's relevance based on the previous assumptions, firstly, we have employed three different formal representations, namely the \textit{I\_in}, the \textit{I\_ex} and the \textit{I\_main}, that have been introduced in Section~\ref{sec:Indexing}. Then, given these three representations, we estimated, for each query, the topical relevance for each of them by using the BM25 model. By doing that, we ended up with three relevance scores that have a different contribution, either positive or negative, to the document's overall relevance. The final step was aggregating those scores by explicitly considering their independent contribution to relevance. That has been achieved by incorporating TOPSIS, a multi-criteria decision analysis method that can provide a final document ranking based on the task-specific characteristics.

Specifically, the implementation of TOPSIS requires several components to be considered:
\begin{itemize}
    \itemsep .2 em
    \item A set of alternatives that have to be ranked, in our case the documents.
    \item A set of considered criteria, with their related performance scores (in our case the three relevance scores), \textit{R\_in}, \textit{R\_ex}, \textit{R\_main} calculated w.r.t. the different document representations using BM25. In order for TOPSIS to be applied, those scores must be monotonically increasing or decreasing and normalized.
    \item A weighting vector associated with the importance of each criterion to the final relevance decision. 
\end{itemize}

\noindent In addition to the components mentioned above, TOPSIS allows to specifically state the contribution, positive or negative, of a criterion to the final overall score. In this particular search task, the \textit{R\_in} and \textit{R\_main} criteria contribute positively, therefore they are \textit{benefit criteria}, while the \textit{R\_ex} contributes negatively, and, as a result, this criterion is a \textit{cost criterion}.

TOPSIS is based on the concept that the chosen alternative, i.e., a document, should have the shortest geometric distance from a positive ideal solution and the longest from a negative ideal solution. In general, the positive ideal solution is a vector consisting of the maximum values of the benefit criteria and the minimum values of the cost criteria. In contrast, the negative ideal solution consists of the maximum values of the cost criteria and the minimum values of the benefit criteria. 

Therefore, given the decision matrix related to a specific query containing the normalized performance (relevance) scores
referred to the considered sections of each document, the weighted average of the matrix can be computed over the three considered criteria, to obtain the final relevance score. Moreover, the positive and negative ideal solutions can be generated s follows: the positive ideal solution is represented by a vector that contains the highest scores of the \textit{R\_in} and \textit{R\_main} criteria and the lowest score of the \textit{R\_ex} criterion, present in the normalized decision matrix, i.e., across all the documents for a given query; the negative ideal solution is the exact opposite.

Finally, for each query-document pair, the euclidean distance of the document from the positive and negative ideal solutions is measured. The documents are then ranked based on the relative closeness of their overall relevance score to the ideal solution. 

\subsection{Neural Approach}
For two of the runs,  namely UNM\_4 and UNM\_5, we implemented a re-ranking approach based on BM25 and Bert. BM25 employs the query representation \textit{Qd} and the document representation \textit{I\_comb*}; on the re-ranking side, Bert for sentences classification was fine-tuned in a low data regime by using collections from previous TREC Biomedical tracks \footnote{\url{http://www.trec-cds.org/}}; and also the collection indicated by the organizers~\cite{DBLP:conf/sigir/KoopmanZ16}. Concerning the document representation, we also used \textit{I\_comb*}. In this way, the combination of query and document does not exceed the limit of tokens of the transformer network. The training procedure was adopted by following previous approaches~\cite{rybinski2020clinical}. 

\section{Experimental Setup}
\label{sec:exp_setup}
Several steps of our methodology required some parameter selection. Therefore, this section presents a few details related to implementation aspects. 

\subsection{Keyword Extraction Method}
To begin with, the keyword extraction method applied to the queries has required the application of a specific model for obtaining both document and word embeddings. Due to its simplicity and speed, we have used the DistilBERT proposed in~\cite{DBLP:journals/corr/abs-1910-01108}. Moreover, the length of the provided queries, in terms of tokens, was smaller than the model's maximum input, and as a result, no action was needed.  

In addition, we needed to set two more parameters related to the diversity between the selected keywords and the size of the new query.
Due to the assumption that each query represents a different patient and that some conditions may be more complicated than others (therefore reflected in longer descriptions) we have modified the original implementation of the EmbedRank++ method. In particular, instead of extracting a constant number of key phrases from each query, we estimate the size of the new query dynamically. The length of the new query is equal to half the size of the initial query; stop-words have been removed. 
Regarding the second parameter, which is related to the diversity between the candidate keywords, it was set equal to 0.5 to capture many different aspects of each query.  

\subsection{Indexing \& Topical Relevance} 
To index the collection and estimate the topical relevance values with respect to each of different document representations (indices), we have employed PyTerrier~\cite{pyterrier2020ictir} and its BM25 implementation, respectively. For preprocessing, we have used the standard PyTerrier's pipeline, i.e., porter-stemming and stopwords removal.  

\subsection{TOPSIS}
As stated in Section~\ref{sec:dt}, TOPSIS associates a weight with each considered criterion based on which each document is evaluated. 
These weights can be either learned from a portion of training data, manually set to specific values, or estimated with some weighting method. 
For the submitted runs R2 and R3, the weights have been set to fixed values equal to 0.33, the mean weight value of the three criteria, thus assuming a same importance for the three criteria. We refer to this model as \textit{TT\_MW} model.

\subsection{Submitted Runs}
Table~\ref{tab:sub_runs} presents the submitted runs along with the corresponding components that have been analyzed in the previous sections.  

\begin{table}[H]
\centering
\def\arraystretch{1.3}
\begin{tabular}{|c|c|c|c|c|}
\hline
Run ID &  Run Name        & Query Representation & Document Index        & Relevance Model \\ \hline
R1      & IKR3\_BSL       & Qk                  & I\_comb               & BM25            \\ \hline
R2      & IKR3\_TT\_MW\_d & Qd                  & I\_in, I\_ex, I\_main & TT\_MW          \\ \hline
R3      & IKR3\_TT\_MW\_k & Qk                  & I\_in, I\_ex, I\_main & TT\_MW          \\ \hline
R4      & UNM\_4          &   Qd                  &     I\_comb*                  &    BM25+BERT             \\ \hline
R5      & UNM\_5          &   Qd                  &       I\_comb*               &         BM25+BERT         \\ \hline
\end{tabular}
\caption{A summary of the components employed in each submitted runs.}
\label{tab:sub_runs}
\end{table}

\section{Results}
\label{sec:evaluation_metrics}
Table~\ref{tab:effectiveness} presents the results obtained from the five submitted runs, while Table~\ref{tab:impr_queries} reports the number of queries that were improved compared to their reported TREC's median performance. Thus, it can be seen that the mean effectiveness of our runs is well-above the averaged TREC's median for three different measures, and also that the submitted runs improve the majority of the queries.  

\begin{table}[!h]
\centering
\def\arraystretch{1.3}
\begin{tabular}{c|c|ccccc}
Measure         & TREC's Median & R1    & R2    & R3  & R4 & R5 \\ \hline
NDCG@10       & .304         & \textbf{.478} & .431 & .449 & .252 & .346\\		
PREC@10       & .161         & .265 & .255 & \textbf{.281} & .152 & .173\\
Reciprocal Rank & .294       & \textbf{.486} & .452 & .459 & .289 & .373\\
NDCG@5  & Not Reported       & \textbf{.504} & .458 & .468 & .278 & .380
\end{tabular}
\caption{Overall comparison with the across query averaged median values, where R1-5 represent the effectiveness obtained by our runs.
}
\label{tab:effectiveness}
\end{table}

\noindent Furthermore, one can observe that the highest PREC@10 value was obtained by the R3 run that involves the TOPSIS model and the Qk representation. Also, the R2 run that exploits the Qd query representation, that is usually more noisy, yields results relatively close to the runs that employ the Qk representation. 

\begin{table}[!htbp]
\centering
\begin{tabular}{c|ccc}
Runs            & R1 & R2 & R3 \\ \hline
NDCG@10         & 63 & 59 & 58 \\
PREC@10         & 63 & 62 & 64 \\
Reciprocal Rank & 53 & 53 & 53
\end{tabular}
\caption{Number of improved queries over the TREC's Median.}
\label{tab:impr_queries}
\end{table}

\noindent Finally, the fact that the associated criteria weights were manually set, instead of learned using a training query set, may have underestimated the effectiveness of the employed decision-theoretic model. 

\section{Conclusions}
In this paper we have presented the results of our participation in the TREC 2021 Clinical Trials Track. Our objective was to investigate the effectiveness of several query representation combined with different retrieval models in this particular search task. 
The analysis of the results confirmed the effectiveness of the query representation approach (run R1). However, we also proved that a decision-theoretic model (run R2), yields similar results compared to a standard retrieval approach that is associated with an ad-hoc query representation (run R1). In addition, the decision-theoretic model was able to improve the system's precision when combined with a better query representation (run R3). Regarding the neural-based approaches, their performance was comparable to the reported TREC's median.

Concluding, it can be seen that using a decision-theoretic model is beneficial for this particular search task. The decision-theoretic model almost met the performance of a standard approach without leveraging the use of the query preprocessing method. At the same time, it outperformed the standard approach in terms of PREC@10 when combined with the employed query preprocessing method. However, the fact that the criteria weights were manually set may have underestimated the model's effectiveness; therefore, further research will be conducted in this direction. 

\section{Acknowledgements}
This work was supported by the EU Horizon 2020 ITN/ETN on Domain Specific Systems for Information Extraction and Retrieval -- DoSSIER (H2020-EU.1.3.1., ID: 860721).

\medskip

\bibliographystyle{unsrt}
\bibliography{trecREF}

\end{document}